\pdfoutput=1

\documentclass[runningheads,final]{llncs}
\usepackage{graphicx}
\usepackage{amsmath}
\usepackage{xcolor}
\usepackage{caption}
\usepackage{subcaption}

\usepackage{booktabs}  

\usepackage[
binary-units,
group-separator={,},
]{siunitx}  
\usepackage{csquotes}  

\usepackage{hyperref}
\usepackage[capitalise]{cleveref}  

\newcommand{\eg}{e.g.\@}
\newcommand{\ie}{i.e.\@}
\newcommand{\cf}{cf.\@}

\newcommand{\versus}{\emph{vs.\@}}

\newcommand{\resp}{resp.\@}
\newcommand{\etal}{et~al.\@}

\makeatletter
\newcommand{\etc}{etc\@ifnextchar.{}{.\@}}
\makeatother

\newcommand{\unit}[1]{\ensuremath{\left[\si{#1}\right]}}
\newcommand{\cluster}[1]{\texttt{#1\relax}}
\newcommand{\linear}[1]{\ensuremath{{#1}_L}}

\begin{document}
\title{%
	Sustaining Performance While Reducing Energy Consumption: A Control Theory Approach
}
\titlerunning{%
	Sustaining Performance While Reducing Energy Consumption
}
%
\author{
Sophie \textsc{Cerf}\inst{1}\orcidID{0000-0003-0122-0796}
\and
Raphaël \textsc{Bleuse}\inst{1}\orcidID{0000-0002-6728-2132}
\and
Valentin \textsc{Reis}\inst{2}
\and
Swann \textsc{Perarnau}\inst{2}\orcidID{0000-0002-1029-0684}
\and
Éric \textsc{Rutten}\inst{1}\orcidID{0000-0001-8696-8212}
}
\authorrunning{S.\@ \textsc{Cerf} \etal}
%
\institute{%
Univ.\@ Grenoble Alpes, Inria, CNRS, Grenoble INP, LIG, F-38000, Grenoble, France \\ \email{first.last@inria.fr}\and
Argonne National Laboratory.
\email{\{vreis,swann\}@anl.gov}
}
\maketitle              
\begin{abstract}

Production high-performance computing systems continue to grow in complexity and size. As applications struggle to
make use of increasingly heterogeneous compute nodes, maintaining high efficiency (performance per watt) for the whole
platform becomes a challenge. Alongside the growing complexity of scientific workloads, this extreme heterogeneity
is also an opportunity: as applications dynamically undergo variations in workload, due to phases or data/compute movement
between devices, one can dynamically adjust power across compute elements to save energy without impacting performance.
With an aim toward an autonomous and dynamic power management strategy for current and future HPC architectures,
this paper explores the use of control theory for the design of a dynamic power regulation method.
Structured as a feedback loop, our approach---which is novel in computing resource management---consists of periodically
monitoring application progress 
and choosing at runtime a suitable power cap for processors. Thanks to
a preliminary offline identification process, we derive a model of the dynamics of the system and a proportional-integral (PI)
controller.
We evaluate our approach on top of an existing resource management framework, the Argo Node Resource Manager, deployed
on several clusters of Grid'5000, using a standard memory-bound HPC benchmark.

\keywords{Power regulation \and HPC systems \and Control theory}
\end{abstract}
\section{Introduction}%
\label{sec:introduction}

Energy efficiency is an ongoing and major concern of production HPC systems. As
we approach exascale, the complexity of these systems increases, leading to
inefficiencies in power allocation schemes across hardware components.
Furthermore, this issue is becoming dynamic in nature: power-performance
variability across identical components in different parts of the system leads
to applications performance issues that must be monitored at runtime in order
to balance power across components. Similarly, application phases can result in
energy inefficiencies. For example, during an application's I/O phase for which
performance is limited by current network capacity, the power allocated to the
processor could be reduced without impact on the application performance.

In this paper we focus on the design of a controller that can dynamically
measure application performance during runtime and reallocate power
accordingly. Our goal is to improve the efficiency of the overall system, with
limited
and controllable impact on application performance. How to design such a
controller is a challenge: while several mechanisms have appeared to
regulate the power consumption of various components (processors, memories,
accelerators), there is no consensus on the most appropriate means to measure
application performance at runtime (as an estimator of total execution time).

At the hardware level, multiple runtime mechanisms have been demonstrated for
regulating power usage. Some of them rely on
DVFS (dynamic voltage and frequency scaling)\cite{AlbersS2011Algorithms}, a frequency and voltage actuator, and build
up a power regulation algorithm\cite{ImesC2015POET,ImesC2019CoPPer}. Another
approach involves using DDCM (dynamic duty cycle modulation) as a power
handle\cite{BhalachandraS2015Using}. More recently, Intel introduced in the
Sandy Bridge microarchitecture RAPL (running average power limit)\cite{DavidH2010RAPL,RotemE2012Power}.  It is an autonomous hardware
solution (i.e., a control loop);
and while the base mechanism behind it remains
not public, RAPL has the benefit of being stable
and widespread across current production
systems.
On the application side, the characterization of its behavior, including phases,
or its compute- or memory-boundedness is more challenging. Indeed, heavy
instrumentation can have an impact on the system itself and result in a change
in behavior, while few mechanisms are available to monitor an application
from the outside. The most versatile such mechanisms, hardware performance counters, require
careful selection of which counters to sample and are not necessarily good
predictors of an application's total execution time (\eg, instructions per seconds
are not enough in memory-bound applications).
Furthermore, existing online solutions rely on simple control
loop designs with limited verifiable properties in terms of stability of the
control or adaptability to application behavior.

In this paper we advocate the use of control theory as a means to formalize
the problem of power regulation under a performance bound, using well-defined
models and solutions to design an autonomous, online control loop with
mathematical guaranties with respect to its robustness. Based on the work of
\textsc{Ramesh}~\etal\cite{RameshS2019Understanding} for an online
application performance estimator (a lightweight heartbeat advertising the
amount of progress towards an internal figure of merit), we design a closed
control loop that acts on the RAPL power cap on recent processors.
The next section provides
background on
measurement and tuning
and 
feedback loop control.
\cref{sec:control-methodology} presents 
the 
workflow of
control theory 
adapted to 
HPC systems.
\cref{sec:modeling-control} details our modeling and control design
, 
validated and discussed in \cref{sec:evaluation}.
In \cref{sec:related-works}
we discuss related works
and 
conclude in \cref{sec:conclusion} with a brief summary.

\section{Background}%
\label{sec:background}


\subsection{Application Measurement and Tuning}

\paragraph{Measuring progress with heartbeats}%
\label{heartbeats}

We follow the work of \textsc{Ramesh} \etal\cite{RameshS2019Understanding} in using
a lightweight instrumentation library that sends a type of application
heartbeat. Based on discussions with application developers and performance
characterization experts, a few places in the application code are identified
as representing significant progress toward the \emph{science} of the
application (or its \emph{figure of merit}). The resulting instrumentation
sends a message on a socket local to the node indicating the amount of progress
performed since the last message. We then derive a \emph{heartrate} from
these messages.

\paragraph{RAPL}

RAPL is a mechanism available on recent Intel processors that allows users to
specify a power cap on available hardware domains, using model-specific
registers or the associated Linux sysfs subsystem. Typically, the
processors make two domains available: the CPU package and a DRAM domain. The
RAPL interface uses two knobs: the power limit and a time window. The internal
controller then guarantees that the average power over the time window is
maintained. This mechanism offers a \emph{sensor} to measure the energy
consumed since the processor was turned on. Consequently, RAPL can be used
to both measure and limit power usage\cite{RountreeB2012Beyond}.

\paragraph{NRM}

The Argo Node Resource Manager\cite{argo-nrm} is an infrastructure for the
design of node-level resource management policies developed as a part of
Argo within the U.S.\@ Department of Energy Exascale Computing Project. It is based on
a daemon process that runs alongside applications and provides to users
a unified interface (through Unix domain sockets) to the various monitoring
and resource controls knobs available on a compute node (RAPL,
performance counters). It also includes libraries to gather application progress,
either through direct lightweight instrumentation or transparently using PMPI.
In this paper all experiments use a Python API that allows users to
bypass internal resource optimization algorithms and implement custom
synchronous control on top of the NRM's bookkeeping of sensor and actuator data.

\subsection{Runtime Self-Adaptation and Feedback Loop Control}
\label{ssec:control-background}


Designing 
feedback loops is
the object of control theory,
which is widespread in all domains of engineering
but
only recently has been scarcely
applied to regulation in computing systems%
\cite{%
HellersteinJL2004Feedback,%
RuttenE2017Feedback%
}.
It provides systems designers
with methodologies to conceive and implement feedback loops with well-mastered 
behavior.
%
%
\label{ControlTheory}
\label{adaptive}
\begin{figure}%
[b]
\centering
\includegraphics[width=0.6\columnwidth]{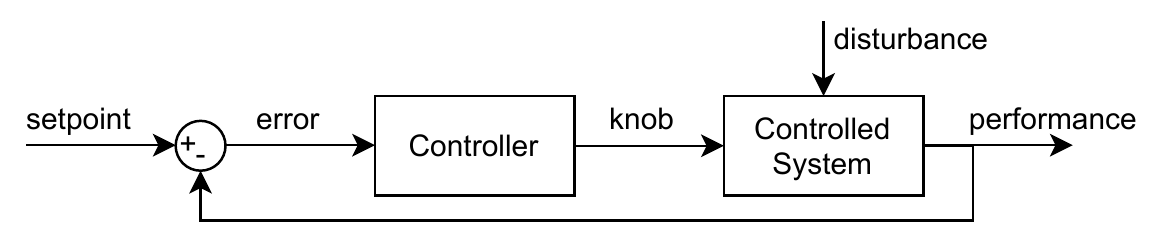}
\caption{Block diagram for a simplistic control loop.}
\label{fig:controlloop}
\end{figure}
In short, the design of control functions is based on an approximated model---since a perfect model is not necessary, nor is it always feasible---of the dynamics of the process to be controlled (\emph{Controlled System} in \cref{fig:controlloop}), in order to derive
controllers with properties including
convergence, avoidance of oscillations,
and mastering of overshoot effects. This process implies the identification of certain variables of interest in the infrastructure:
	the \emph{performance}, which is a measurable indicator of the state of the system;
	the \emph{setpoint}, the objective value to which we want to drive the \emph{performance};
	the \emph{knob}, the action through which the \emph{performance} can be modified or regulated to match the \emph{setpoint} value;
	the \emph{error}, the difference between the \emph{setpoint} value and the system's \emph{performance}, as a measure of deviation; and
	the \emph{disturbance}, an external agent that affects the system's dynamics in an unpredictable way.

Through the knowledge of a system's dynamical model, the approaches of control theory can deliver management strategies that define how to adjust system's knobs to get the desired performance: \ie, to control the state of the system.
Traditional control solutions include proportional-integral-derivative
controllers\cite{LevineWSControl} where
the value of the command to be executed is given by an equation with three terms:
\(P\) proportional to the \emph{error},
\(I\) involving an integration over time of the error past values (\ie, a memory-like effect),
and \(D\) based on its current \enquote{speed} or rate of change, which takes into account possible future trends.


\section{Control Methodology for Runtime Adaptation of Power}
\label{sec:control-methodology}

\begin{figure}
    \centering
    \includegraphics[width=\linewidth]{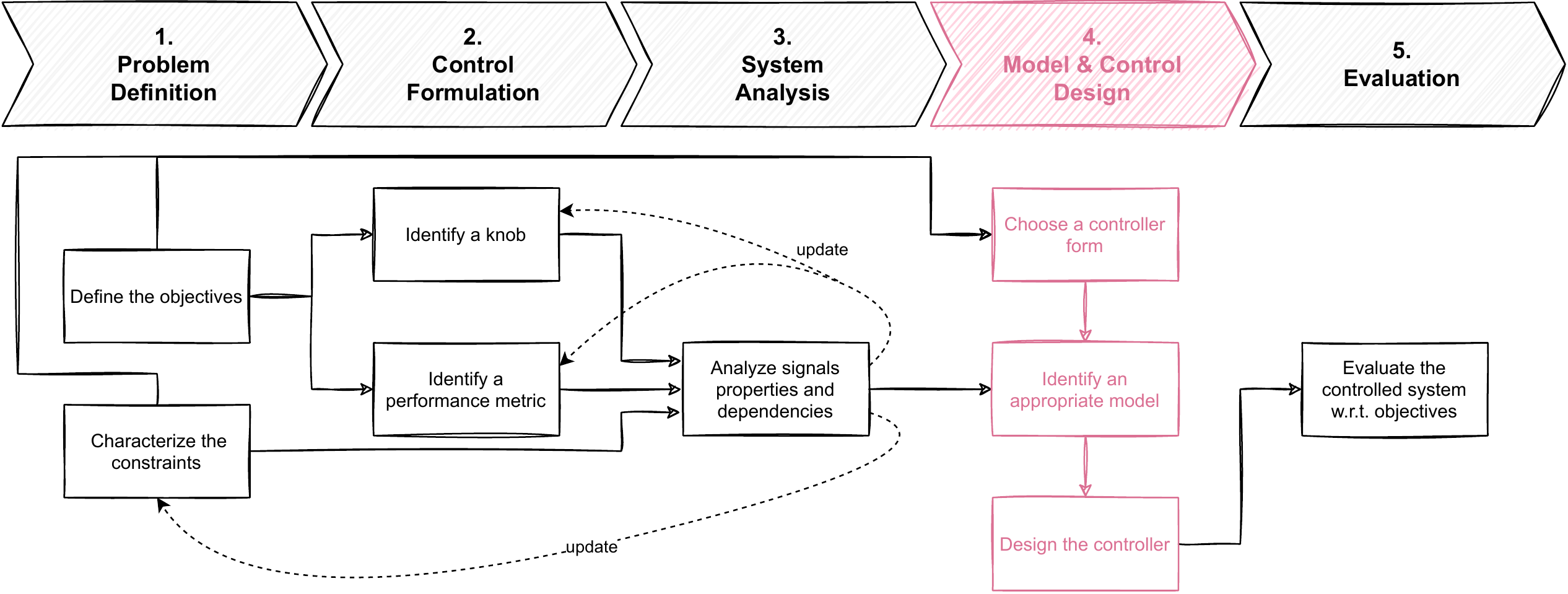}
    \caption{Methodological steps in using control theory for HPC systems. Step 4 (highlighted in pink) can be done independently by control experts.}
    \label{fig:control-methodo}
\end{figure}

The use of control theory tooling for computing systems is recent, especially in HPC\cite{StahlE2018Towards}.
The community still lacks well-defined and accepted models.
Needed, therefore, is a methodology to identify and construct models upon which controllers can be designed.
%
 \cref{fig:control-methodo} depicts the method
 used for this work.
 It adapts the traditional workflow of control theory\cite{FilieriA2017Control},
by explicitly considering
the upstream work of defining the problem and translating it to a control-friendly formulation
in
five main steps:
%
\begin{description}
	\item[(1) Problem definition]
		First, the objectives have to be settled and the challenging characteristics and constraints identified. Here the problem definition naturally results from \cref{sec:introduction,sec:background}.
		It consists of sustaining the application execution time while reducing energy usage as much as possible.
		This is challenging given the following constraints: applications have unpredictable phases with exogenous limits on progress, progress metrics are application-specific, processor characteristics and performance are various, power actuators have a limited accuracy and are distributed on all packages, and thermal considerations induce nonlinearities.
%
	\item[(2) Control formulation]
		The objectives are then translated into suitable control signals: these are knobs that enable one to act on the system at runtime and performance metric(s) to monitor the system's behavior.
		They need to be measurable or computable on the fly.
		For example, the RAPL powercap value and the application progress are such signals.
	\item[(3) System analysis]
		The knobs are then analyzed to assess the impact of both their levels and the change of their levels on the performance signals during the execution time (\cf\ \cref{ssec:system-analysis}).
		The control formulation may be updated after this analysis to ease the control development (select adequate sampling time, modify signals with logarithmic scales, offsets).
		New constraints may be highlighted, such as the number of packages in a node.
	\item[(4) Model and control design]
		Once the relationships between signals and the system's behavior have been identified, the control theory toolbox may be used\cite{LevineWSControl}.
		According to the objectives and the identified system characteristics challenging them, a controller form is chosen, a PI controller in this case. An adequate model is identified from experimental data, such as a first-order dynamical model (see \cref{ssec:modeling}), and then used to design and tune the controller, following the pole placement method here (see \cref{ssec:control}).
	\item[(5) Evaluation]
		Eventually, the controlled system is evaluated with respect to the objectives (\cf\ \cref{sec:evaluation}).
		In this paper we are interested in evaluating the energy reduction and execution time increase according to the maximal allowed degradation given to the controller.
\end{description}

\section{Model and Controller Development}
\label{sec:modeling-control}

We defined the problem under study 
in \cref{sec:introduction,sec:background}.
Following the method described in \cref{sec:control-methodology}, we now focus on formulating it as a control problem, analyzing the system to derive a model and design a controller (Steps 2 to 4).
As depicted in \cref{fig:control-methodo}, the early steps are refined in an iterative process with respect to the analysis results.
Since the analysis of the system requires observing the behavior of the system, we first describe the experimental setup.

\subsection{Experimental Setup}
\label{ssec:setup}

\paragraph{Platform}

All experiments were conducted on the Grid'5000 testbed.
We ran the experiments on nodes of three different clusters: \cluster{gros}, \cluster{dahu}, and \cluster{yeti}.
These clusters were chosen because their nodes have modern Intel CPUs and a varying number of sockets.
We list in \cref{tab:g5k-clusters} the main characteristics of the clusters.
The exact specifications of the clusters are available on the Grid'5000 wiki.%
\footnote{\url{https://www.grid5000.fr/w/Hardware} with reference API version \texttt{\href{https://github.com/grid5000/reference-repository/commit/9925e0598}{9925e0598}}.}

\begin{table}
[t]
\setlength{\tabcolsep}{1em}
\begin{center}
\begin{tabular}{*{5}{l}}
	\toprule
	Cluster & CPU & Cores/CPU & Sockets & RAM \unit{\gibi\byte} \\
	\midrule
	\cluster{gros} & Xeon Gold 5220 & 18 & 1 & 96 \\
	\cluster{dahu} & Xeon Gold 6130 & 16 & 2 & 192 \\
	\cluster{yeti} & Xeon Gold 6130 & 16 & 4 & 768 \\
	\bottomrule
\end{tabular}
\end{center}
\caption{Hardware characteristics of Grid'5000 clusters used for the experiments.}
\label{tab:g5k-clusters}
\end{table}

\paragraph{Software stack}

All experiments ran on a deployed environment with a custom image.
The deployed environment is a minimal GNU/Linux Debian \texttt{10.7} \enquote{buster} with kernel \texttt{4.19.0-13-amd64}.
The management of applications and resources was implemented within the Argo
NRM, a resource management framework developed at Argonne National Laboratory.
We used the version tagged as \texttt{expe-0.6} for this work.%
\footnote{Available at \url{https://xgitlab.cels.anl.gov/argo/hnrm}.}
NRM and benchmarks are packaged with the Nix functional package manager\cite{DolstraE2004Nix}: we rely on a multiuser installation of Nix version \texttt{2.3.10}.

\paragraph{Benchmark}

All experiments involved execution of the STREAM benchmark\cite{McCalpinJD1995Memory}.
STREAM is chosen as it is representative of memory-bound phases of applications and shows a stable behavior.
STREAM is also easy to modify into an iterative application, which allows computation of the progress metric by reporting heartbeats.
We used version \texttt{5.10} with a problem size set to \num{33 554 432} and \SI{10 000}{iterations}, further adapted to run in a way that progress can be tracked: its 4 kernels ran a configurable number of times in a loop, with a heartbeat being reported to the NRM each time the loop completed (after one run of the four kernels).

\paragraph{Characterization \versus\ Evaluation setup}

Although the hardware and software stack remain the same, we need to distinguish characterization and evaluation experiments.
For the analysis of the system (characterization), we observe the behavior of the system, and the resource manager follows a predefined plan.
This contrasts with the evaluation setup, where the resource manager reacts to the system's behavior.
From the control theory perspective, the former setup is an open-loop system while the latter is a closed-loop system.

\subsection{Control Formulation}
\label{ssec:control-formulation}

The power actuator is RAPL's power limit denoted \(\text{pcap}(t_i)\).
To define a progress metric, 
we aggregate the heartbeats that the application generates at times \(t_k\) (see \cref{heartbeats}) into a signal synchronized with the power actuator.
The progress metric at \(t_i\) is formally defined as the median of the heartbeats arrival frequencies since the last sampling time \(t_{i-1}\):
\begin{equation}
	\text{progress}(t_i) = \underset{ \forall k,\,t_k \in \left[t_{i-1}, t_i\right[}{\operatorname{median}} \left( \frac{1}{t_k - t_{k-1}} \right)
	\label{eq:progress-definition}
\end{equation}

A central tendency indicator, 
in particular the median, was selected to be robust to extreme values so as to provide the controller with a smooth signal.
The primary performance objective is based on the application's execution time, while for control purposes a runtime metric is needed.

Before further control development, let us ensure the correlation between the two.
We compute the Pearson correlation coefficient\cite{MontgomeryDC2018Applied} between the progress metric and execution time.
The Pearson value is respectively \(0.97\), \(0.80\), and \(0.80\) on \cluster{gros}, \cluster{dahu}, and \cluster{yeti} clusters when computed by using static characterization data (constant powercap over the whole benchmark duration; see \cref{sssec:static-model}). 
The high correlation results validate the progress definition choice, with a notably strong correlation on the 1-socket cluster.

\subsection{System Analysis}
\label{ssec:system-analysis}


The analysis phase assesses
the trustworthiness of
the power actuator and progress sensor
and
measures how powercap levels impact progress.

\begin{figure*}
[!ht]
     \centering
     \begin{subfigure}[b]{0.32\textwidth}
         \centering
         \includegraphics[width=1.1\textwidth]{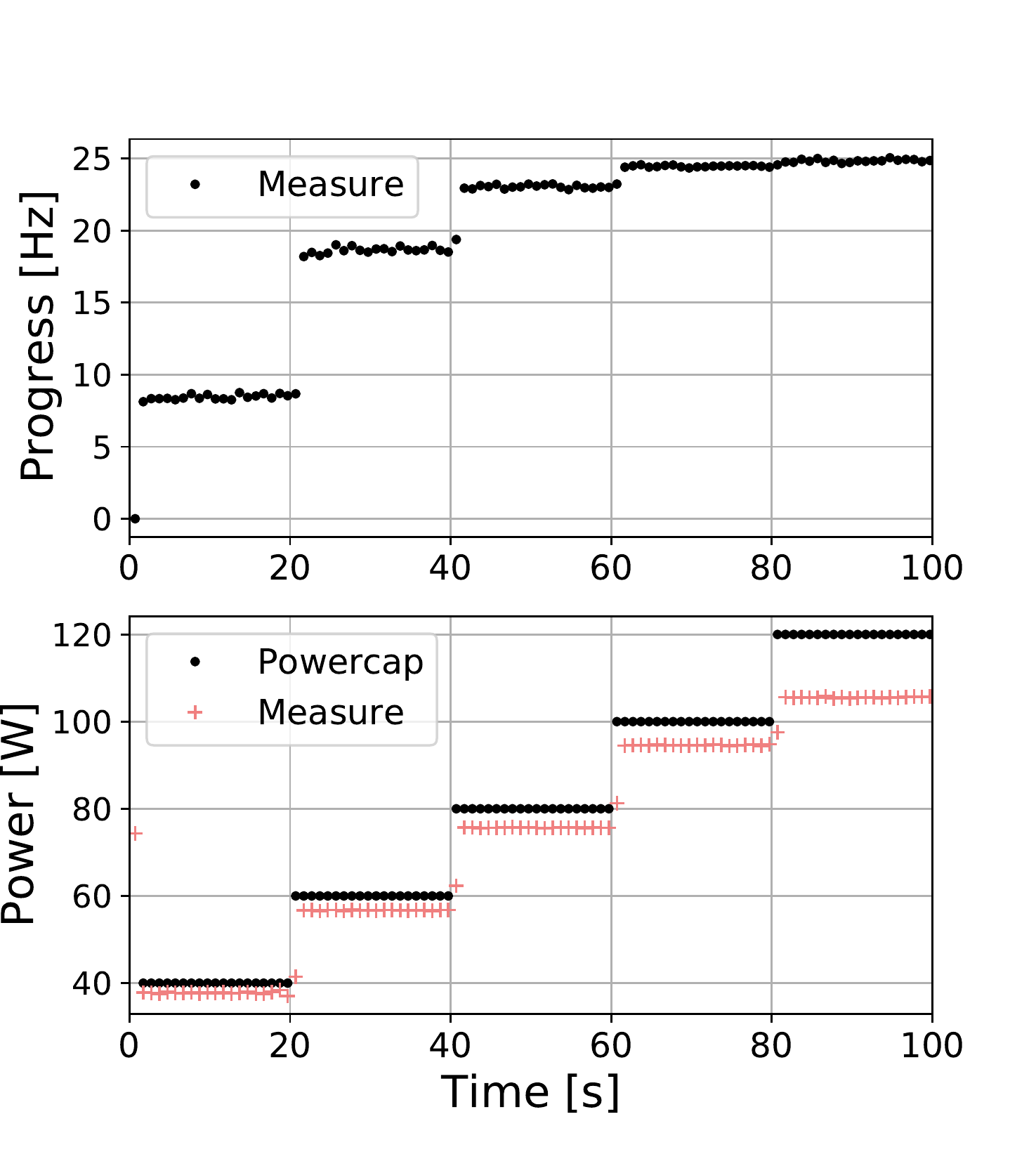}
         \caption{\cluster{gros} cluster}
         \label{fig:identification-stairs-gros}
     \end{subfigure}
     \hfill
     \begin{subfigure}[b]{0.32\textwidth}
         \centering
         \includegraphics[width=1.1\textwidth]{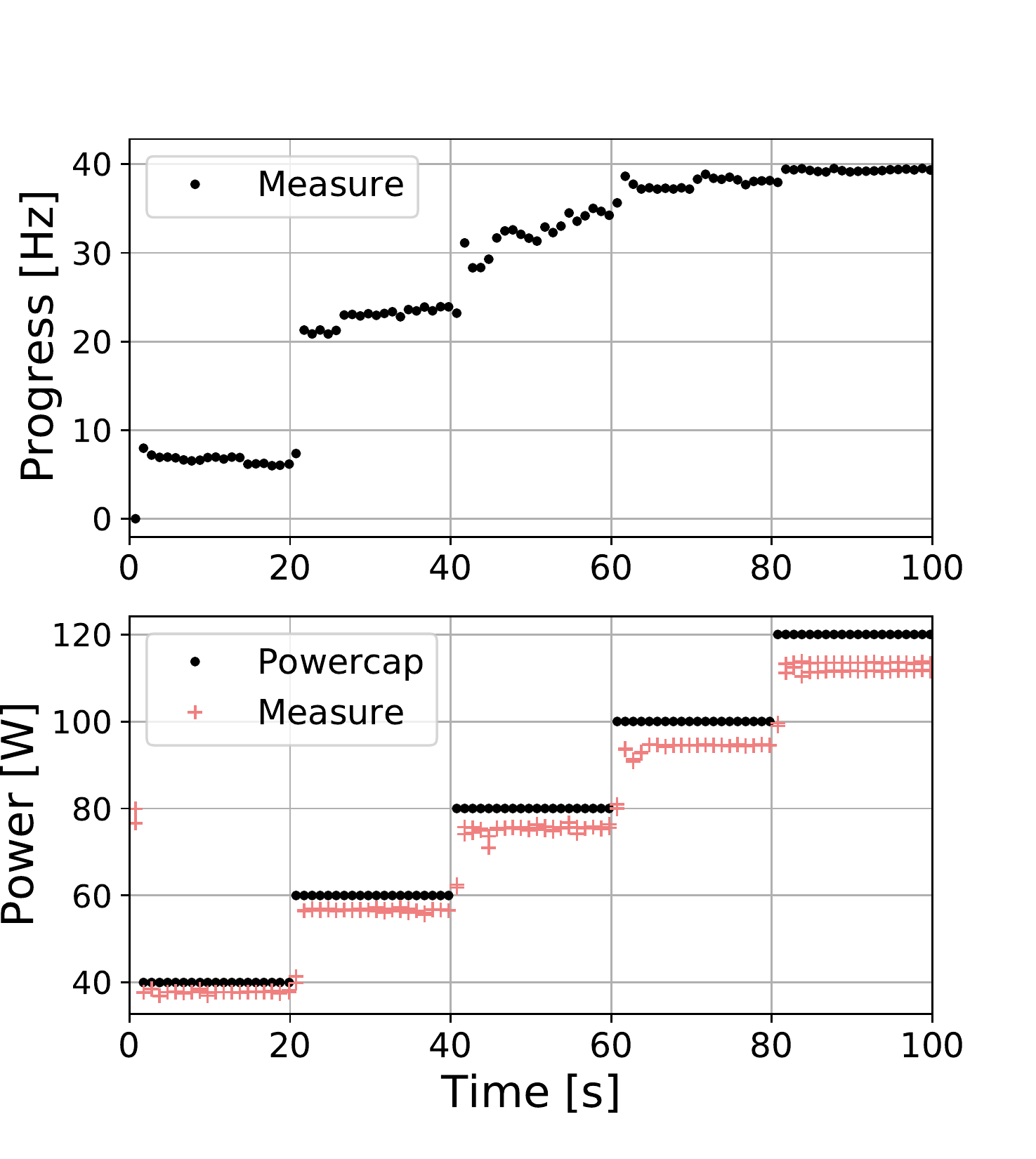}
         \caption{\cluster{dahu} cluster}
         \label{fig:identification-stairs-dahu}
     \end{subfigure}
     \hfill
     \begin{subfigure}[b]{0.32\textwidth}
         \centering
         \includegraphics[width=1.1\textwidth]{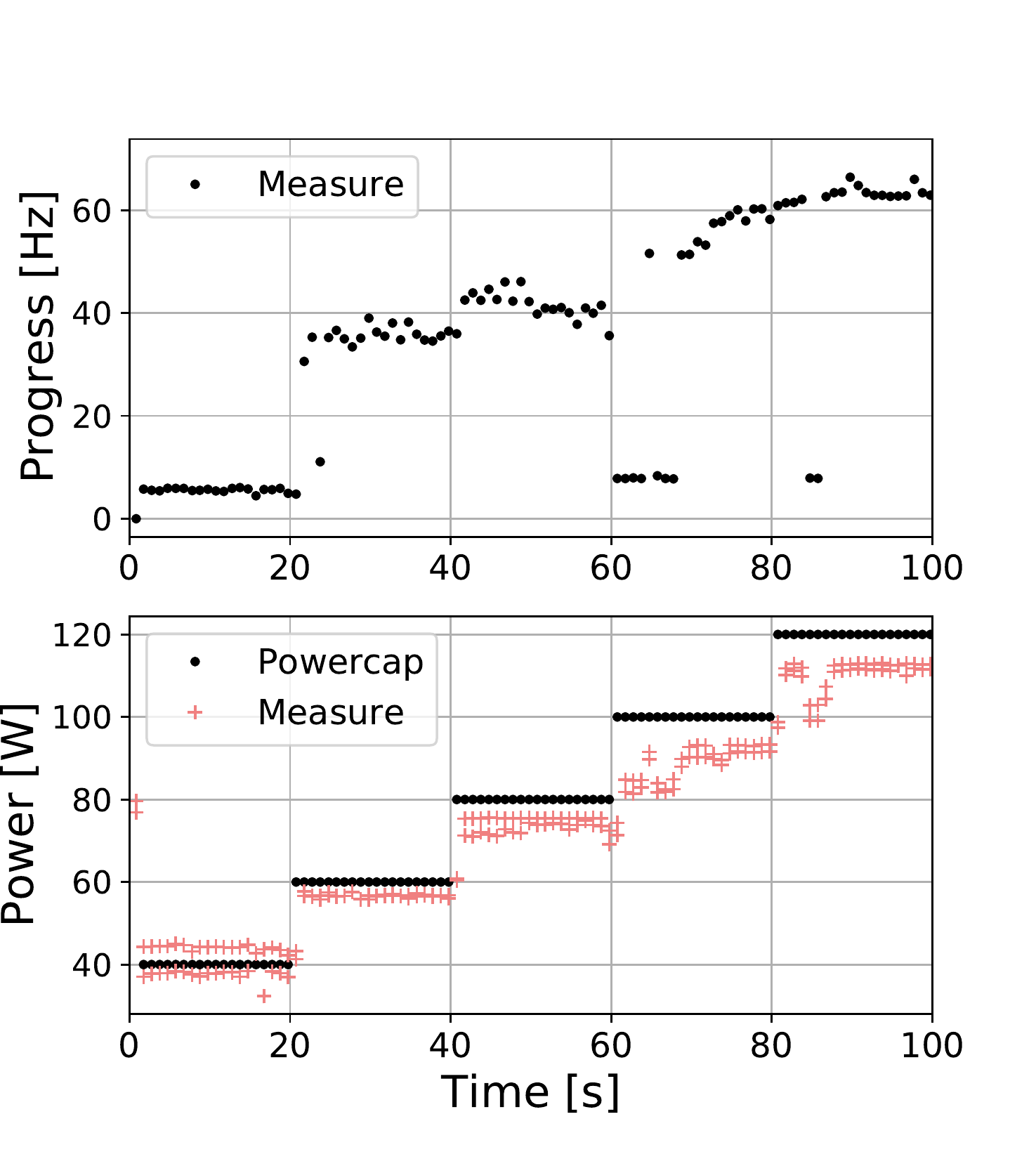}
         \caption{\cluster{yeti} cluster}
         \label{fig:identification-stairs-yeti}
     \end{subfigure}
     \hfill
        \caption{%
            Impact of power changes on progress: the time perspective.
            Each sub-figure depicts a single representative execution.
        }
        \label{fig:identification-stairs}
\end{figure*}

During the benchmark execution, the powercap is gradually increased by steps of \SI{20}{\watt} on the clusters' reasonable power range (\ie, from \SIrange{40}{120}{\watt}), and the progress is measured; see \cref{fig:identification-stairs}. First, we see that the measured power never corresponds to the requested level and that the error increases with the powercap value. The RAPL powercap actuator accuracy is poor\cite{DesrochersS2016Validation} and will have to be taken into account.
The progress variations follow the power ones, whereas the higher the power level, the less a power increase impacts the progress. This highlights the nonlinearity of the power-to-progress dynamical system, with a saturation effect at high power values.
The saturation results from the memory-boundedness of the application: at high power levels, the processor power is not limiting performance as memory is.
The power level at which the saturation appears is related to the processors' thermal design power.
We also note that the more packages there are in the cluster, the noisier the progress.
Additionally, \cref{fig:identification-stairs-yeti} illustrates that the progress is impacted by external factors, since in this run the power signal in itself does not explain the progress plateau from \SIrange{20}{60}{\second} and the drop to \SI{10}{\hertz} at \SI{80}{\second}.
This behavior is further discussed in \cref{ssec:discussion}.

In summary, this analysis shows that soundly tuning the power level enables application progress to be maintained while reducing the energy footprint. This highlights the need for runtime feedback to cope with external factors.
Since all clusters show similar behavior, a common controller can be designed. Cluster-specific modeling will further enable leveraging of its parameters.


\begin{figure}%
[t]
     \centering
     \begin{subfigure}[b]{0.49\textwidth}
         \centering
         \includegraphics[width=1.1\textwidth]{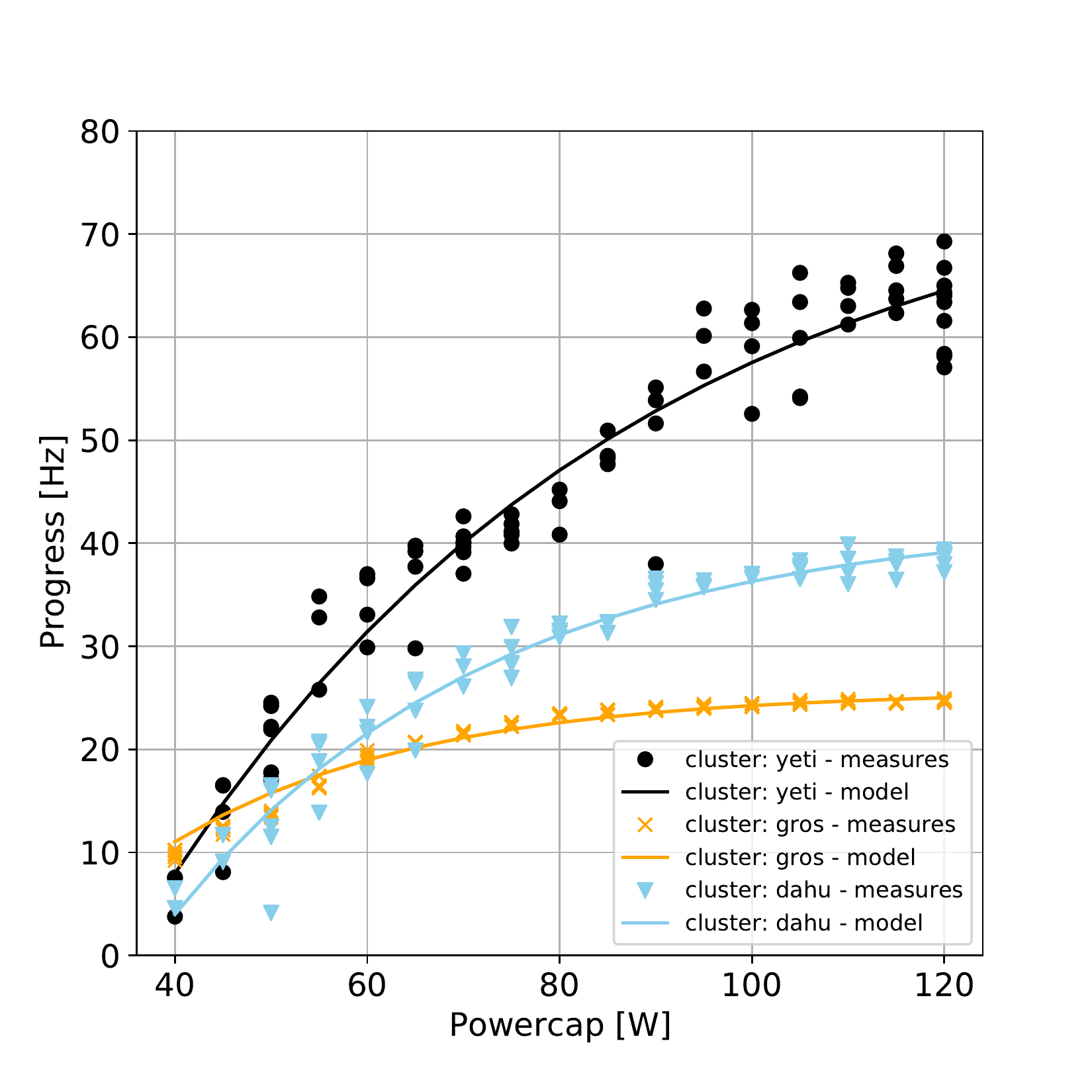}
         \caption{From powercap to progress.}
         \label{fig:sc-heartbeat-frequency-model}
     \end{subfigure}
     \hfill
     \begin{subfigure}[b]{0.49\textwidth}
         \centering
         \includegraphics[width=1.1\textwidth]{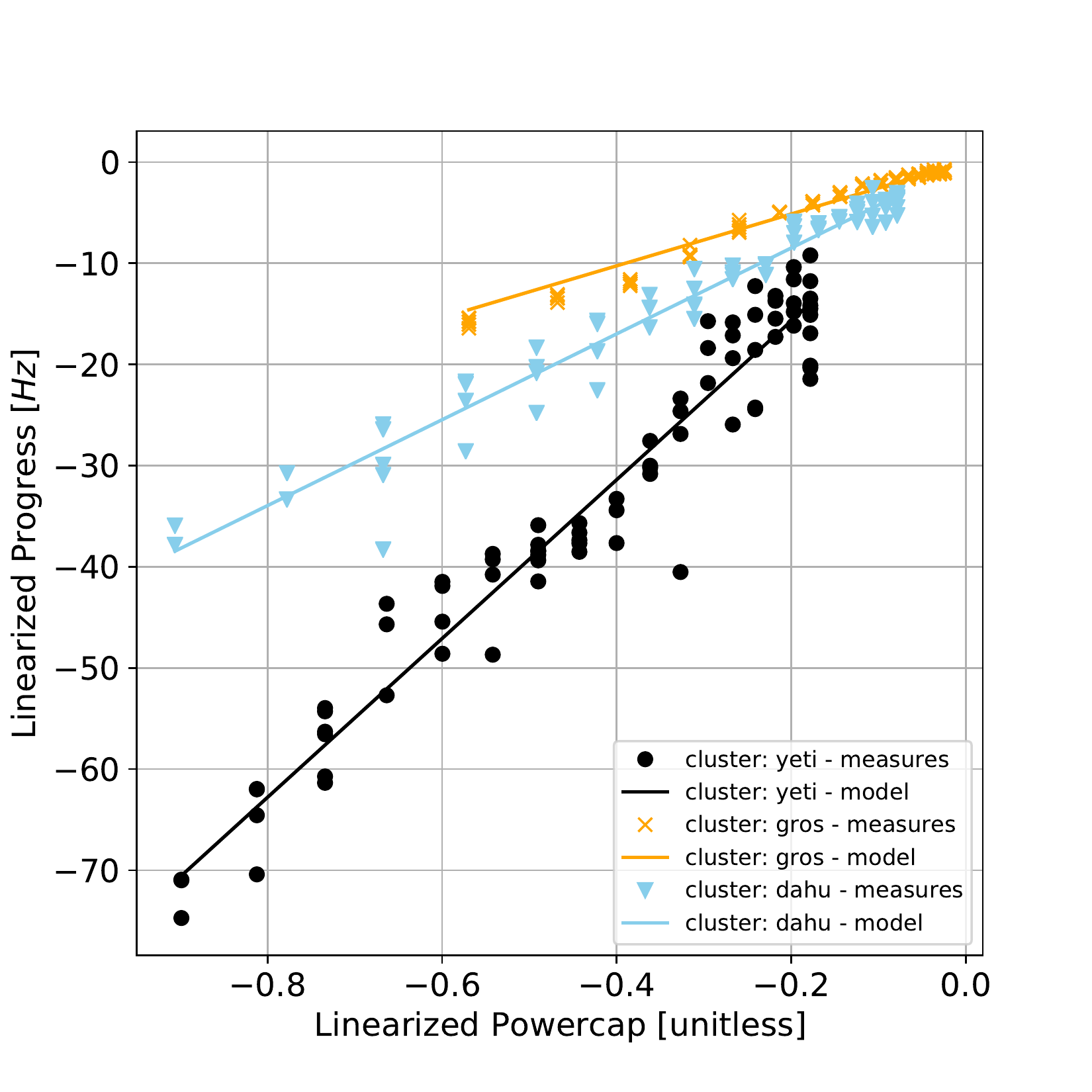}
         \caption{Linearized model.}
         \label{fig:sc-linearized-model}
     \end{subfigure}
     \hfill
        \caption{%
            Static characteristic: modeling of time-averaged behavior.
            Each point depicts a single execution.
        }
        \label{fig:sc-model}
\end{figure}

\subsection{Modeling}
\label{ssec:modeling}

A system model---namely, a set of equations linking the power and progress---is
a first 
step toward a controller taking sound actions.
This step needs to be performed only once per cluster to measure model parameters 
to tune the controller.

\subsubsection{Static characteristic: averaged behavior.}
\label{sssec:static-model}

The time-averaged relation between power and progress is first modeled in a so-called static characterization, in the sense that it considers stabilized situations.
In \cref{fig:sc-heartbeat-frequency-model}, each data point corresponds to an entire benchmark execution where a constant powercap is applied (at the level reported on \(x\)-axis) and for which the progress signal is averaged (\(y\)-axis).
We depict with different colors and markers the measures for the three clusters we use in this work.
For each cluster, at least \SI{68}{experiments} were run.
Note that the curves are flattening, indicating the nonlinear behavior and saturation at large power previously identified.

Based on those experiments, 
a static model, linking the \emph{time-stabilized} powercap to the progress 
is:
%
\(
\text{progress} = K_L \left( 1 - e^{ - \alpha \left( a \cdot \text{pcap} + b - \beta \right)} \right)
    \label{eq:sc-model}
\)
%
where \(a\) and \(b\) parameters represent RAPL actuator accuracy on the cluster (slope and offset, \resp: \(\text{power} = a \cdot \text{pcap} + b\)), \(\alpha\) and \(\beta\) characterize the benchmark-dependent power-to-progress profile, and \(K_L\) is the \emph{linear gain} being both benchmark and cluster specific. The effective values of these parameters can be automatically found by using nonlinear least squares; they are reported in \cref{tab:model-params}.
The solid lines in \cref{fig:sc-heartbeat-frequency-model} illustrate our model, which shows good accuracy (\(0.83<R^2<0.95\)).

\paragraph{Control formulation update.}
Given the nonlinearity of \cref{eq:sc-model}, we simplify the control by linearizing the powercap and progress signals; see \cref{fig:sc-linearized-model}.
The linearized version of a signal~\(\star\) is denoted by \linear{\star}:
\begin{equation}
	\linear{\text{pcap}} = - e^{ - \alpha \left( a \cdot \text{pcap} + b - \beta \right)}%
	\quad%
	;%
	\quad%
	\linear{\text{progress}} = \text{progress} - K_L
\label{eq:pcap-progress-linear}
\end{equation}

%
%

\subsubsection{Dynamic perspective: modeling impact of power variations.}
\label{sssec:dynamic-model}

We now characterize the dynamics, that is, the effects of a change of the power level on the progress signal \emph{over time}.
First, the model form is selected. As a compromise between accuracy and simplicity, we opt for a first-order model\cite{HellersteinJL2004Feedback}. It means that the prediction of the progress requires only one previous value of the measured progress and the enforced powercap level:
\(
\linear{\text{progress}}(t_{i+1}) = f \left( \linear{\text{progress}}(t_i), \linear{\text{pcap}}(t_i) \right).
\)
%
Control theory provides a first-order model formulation based on the static characteristic gain \(K_L\) and on a time constant \(\tau\) characterizing the transient behavior:
\begin{equation}
\linear{\text{progress}}(t_{i+1}) = \frac{K_L \Delta t_i}{\Delta t_i + \tau} \cdot \linear{\text{pcap}}(t_i) + \frac{\tau}{\Delta t_i + \tau} \cdot \linear{\text{progress}}(t_i)
\label{eq:model-linear}
\end{equation}
where \(\Delta t_i = t_{i+1} - t_i\). 
Based on experimental
data
, 
\(\tau=\SI{1/3}{\hertz}\) for all clusters.


\subsection{Control Design}
\label{ssec:control}

The control objective is given as a degradation factor \( \epsilon \), that is, the tolerable loss of performance. 
The controller translates \( \epsilon \) in a progress setpoint to track using the maximum progress (\( \text{progress}_{\text{max}} \)) estimated by using \cref{eq:sc-model} with the cluster maximal power.
A feedback PI controller is developed (see \cref{ssec:control-background}), setting the powercap proportionally to the progress error \(e(t_i)=(1-\epsilon)\cdot\text{progress}_{\text{max}} - \text{progress}(t_i)\) and to the integral of this error (see \cref{ssec:control-background}):
\begin{equation}
\begin{split}
\linear{\text{pcap}}(t_i) &= \left(K_I \Delta t_i + K_P\right) \cdot e(t_i) - K_P \cdot e(t_{i-1}) + \linear{\text{pcap}}(t_{i-1})
\end{split}
\end{equation}

The 
parameters \(K_P\) and \(K_I\) are 
based both on the model parameters \(K_L\) and \( \tau \) and on a tunable parameter \( \tau_{\text{obj}} \): \(K_P = {\tau}/{\left(K_L \cdot \tau_{\text{obj}}\right)}\) and \(K_I = {1}/{\left(K_L \cdot \tau_{\text{obj}}\right)}\),
with \(\tau_{\text{obj}}\) defining the desired dynamical behavior of the controlled system\cite{AstromKJ1995PID}. 
The controller is chosen to be nonaggressive
,
tuned with \( \tau_{\text{obj}} = \SI{10}{\second} > 10 \tau \).
The 
powercap is computed from its linearized value by using \cref{eq:pcap-progress-linear}.


\section{Evaluation}
\label{sec:evaluation}

The experimental setup has been described in \cref{ssec:setup}.
We evaluate 
here the performance of the model and controller designed in the preceding section using the memory-bound STREAM benchmark run on three clusters with varying number of sockets.

We recapitulate in \cref{tab:model-params} the values of the model and controller parameters.
The model parameters \(a\), \(b\), \(\alpha\), and \(\beta\) have been fitted for each cluster with the static characterization experiments (\cf\ \cref{fig:sc-model}).
The model parameters \(K_L\) and \(\tau\) and the  controller parameter \(\tau_{\text{obj}}\) have been chosen with respect to the system's dynamic (\cf\ \cref{fig:identification-stairs}).
These values have been used for the evaluation campaign.

\begin{table}
[t]
\setlength{\tabcolsep}{1em}
\begin{center}
\begin{tabular}{lcc|*{3}{c}}
	\toprule
	Description   & Notation                & Unit             & \cluster{gros} & \cluster{dahu} & \cluster{yeti} \\
	\midrule
	RAPL slope    & \(a\)                   & \unit{1}         & 0.83           & 0.94           & 0.89 \\
	RAPL offset   & \(b\)                   & \unit{\watt}     & 7.07           & 0.17           & 2.91 \\
	              & \(\alpha\)              & \unit{\per\watt} & 0.047          & 0.032          & 0.023 \\
	power offset  & \(\beta\)               & \unit{\watt}     & 28.5           & 34.8           & 33.7 \\
	linear gain   & \(K_L\)                 & \unit{\hertz}    & 25.6           & 42.4           & 78.5 \\
	time constant & \(\tau\)                & \unit{\second}   & 1/3            & 1/3            & 1/3 \\
	\midrule
	              & \(\tau_{\text{obj}}\)   & \unit{\second}   & 10             & 10             & 10 \\
	\bottomrule
\end{tabular}
\end{center}
\caption{Model and controller parameters for each cluster.}
\label{tab:model-params}
\end{table}

\subsection{Measure of the Model Accuracy}

\begin{figure}
[ht]
     \centering
     \begin{subfigure}[b]{0.32\textwidth}
        \begin{flushright}
         \includegraphics[width=1.13\textwidth]{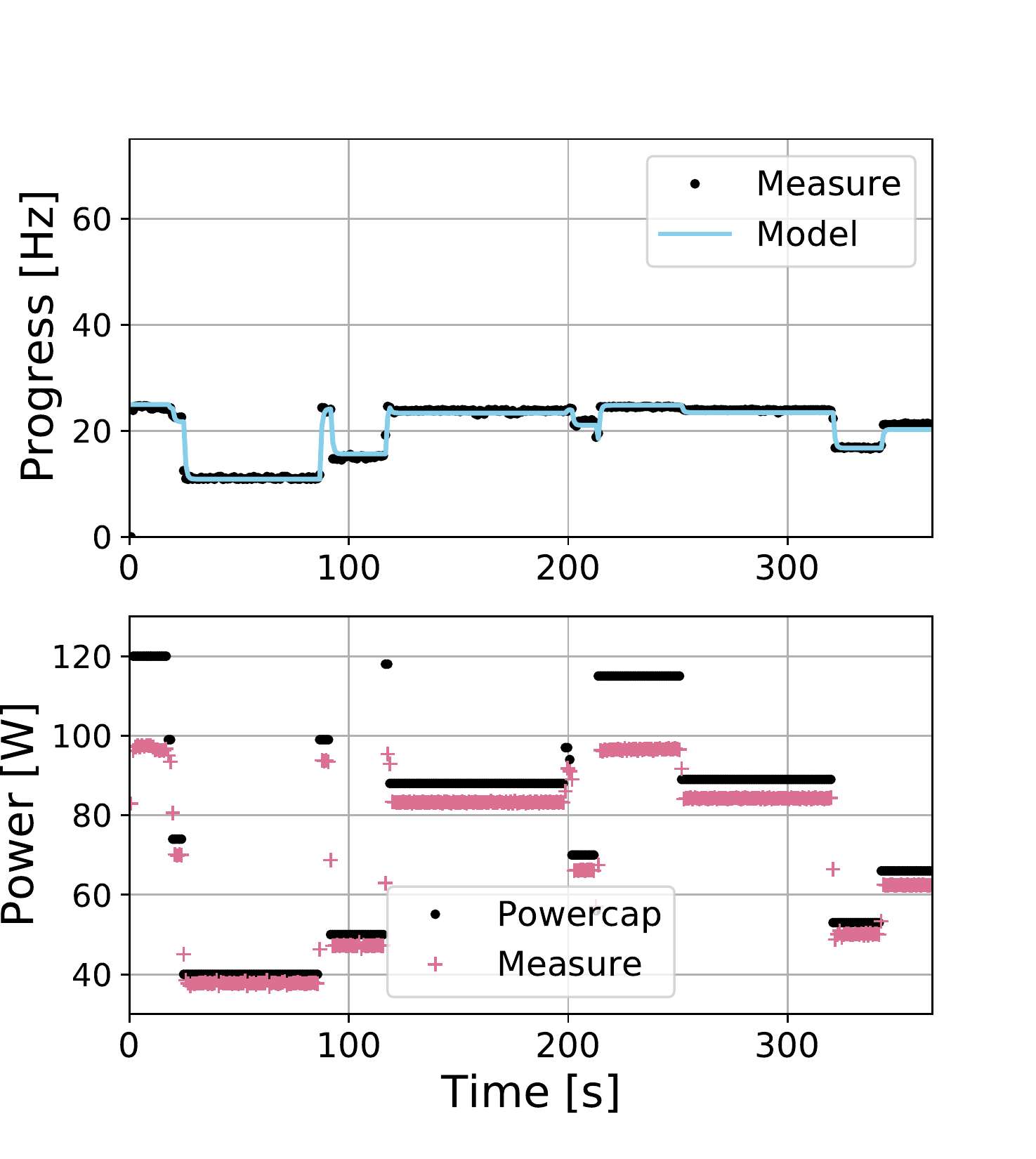}
         \caption{\cluster{gros} cluster}
         \label{fig:identification-random-model-gros}
         \end{flushright}
     \end{subfigure}
     \hfill
     \begin{subfigure}[b]{0.32\textwidth}
         \centering
         \includegraphics[width=1.13\textwidth]{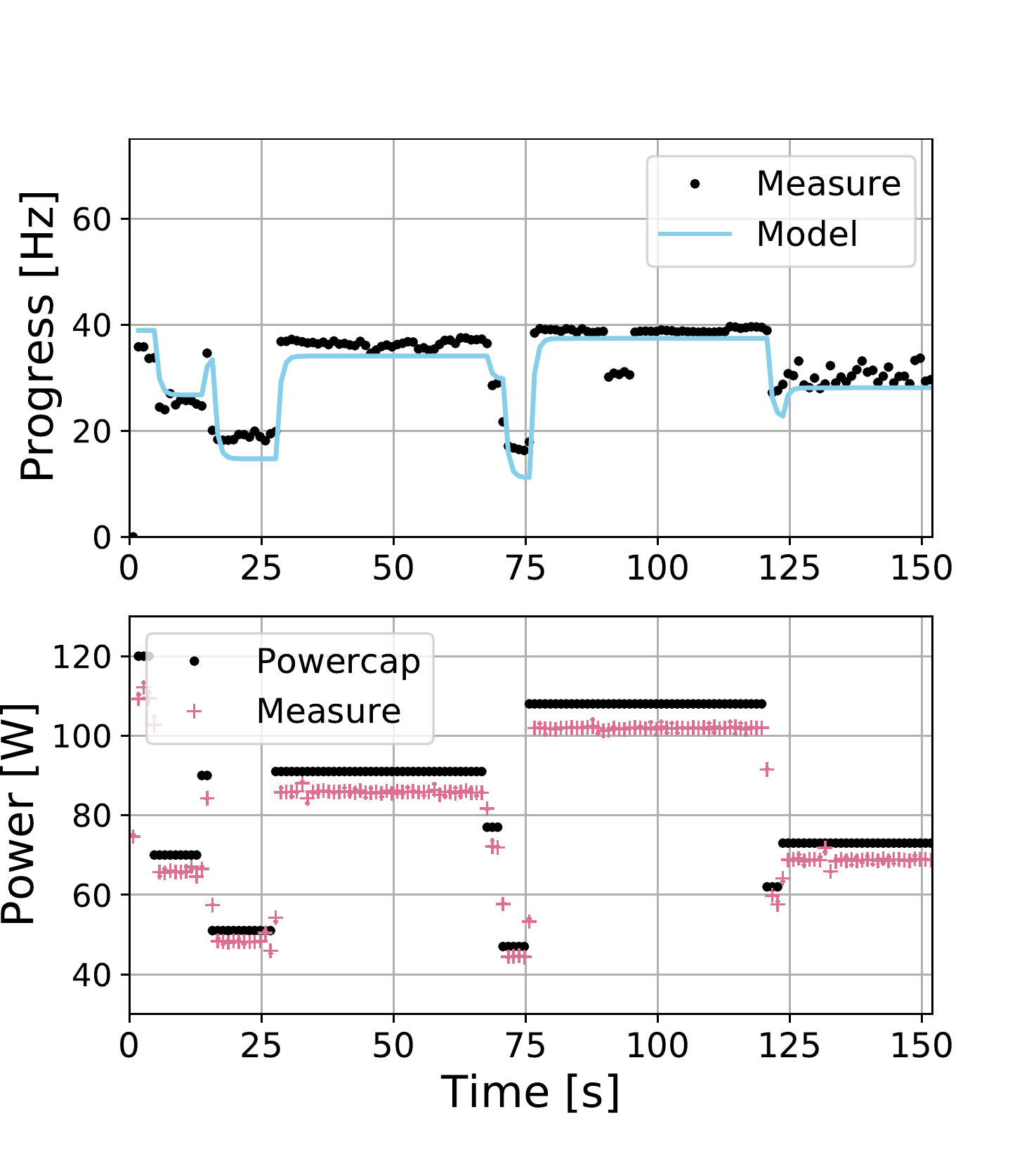}
         \caption{\cluster{dahu} cluster}
         \label{fig:identification-stairs-model-dahu}
     \end{subfigure}
     \hfill
     \begin{subfigure}[b]{0.32\textwidth}
         \centering
         \includegraphics[width=1.13\textwidth]{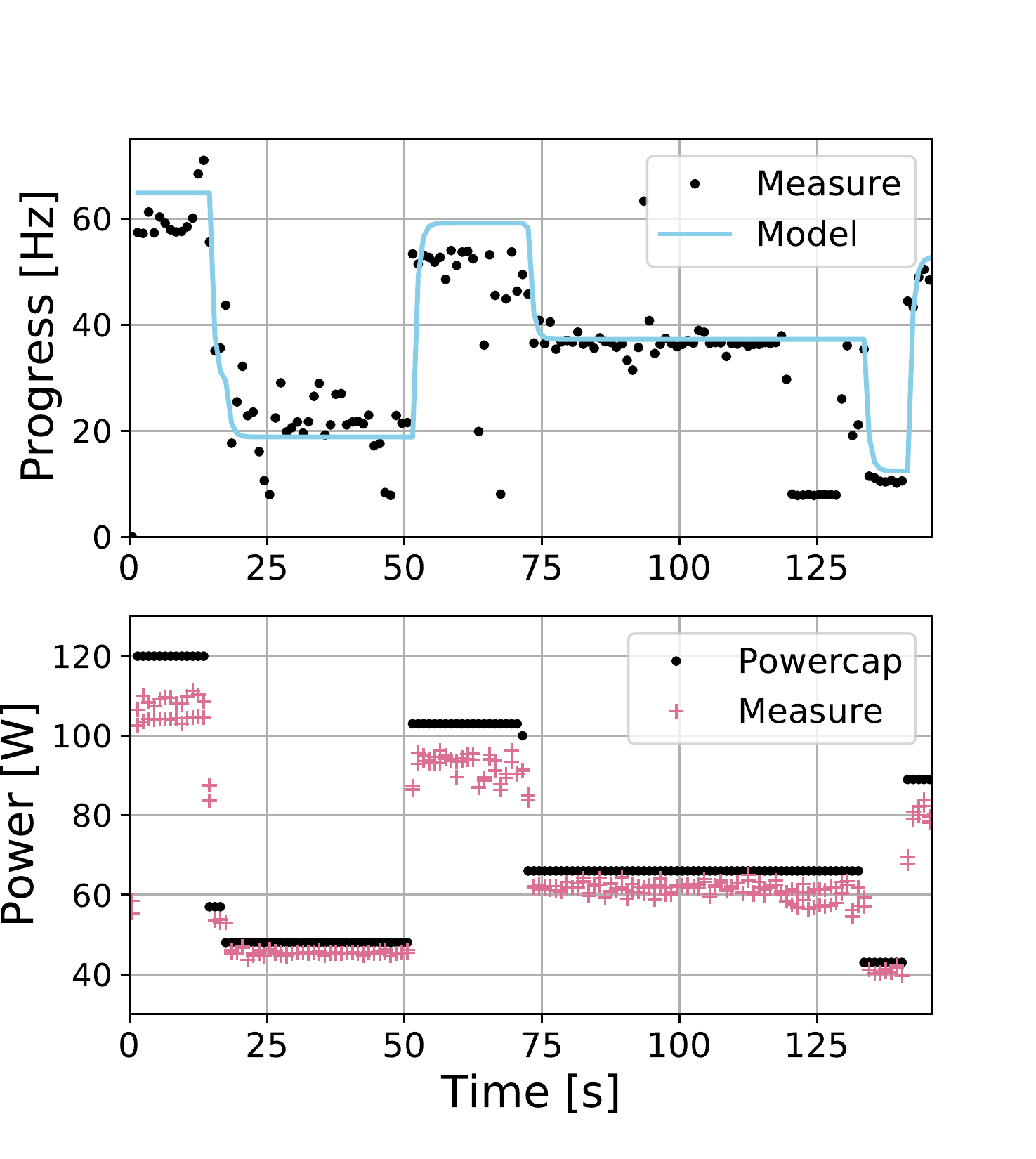}
         \caption{\cluster{yeti} cluster}
         \label{fig:identification-stairs-model-yeti}
     \end{subfigure}
     \hfill
        \caption{%
            Modeling the time dynamics.
            Each sub-figure depicts a single representative execution.
        }
        \label{fig:identification-stairs-model}
\end{figure}

The presented model is not intended to perform predictions: it is used only for the controller tuning. However, we take a brief look at its accuracy. To do so, a random powercap signal is applied, with varying magnitude (from \SIrange{40}{120}{\watt}) and frequency (from \SIrange{e-2}{1}{\hertz}), and the benchmark progress is measured.
For each cluster, at least 20 of such identification experiments were run.
\Cref{fig:identification-stairs-model} illustrates a single execution for each cluster, with the progress measure and its modeled value through time on the top plots and the powercap and power measures on the bottom ones. Visually, the modeling is fairly accurate, and the fewer the sockets, the less noisy the progress metric and the better the modeling.
The average error is close to zero for all clusters.
Nevertheless, our model performs better on clusters with few sockets (narrow distribution and short extrema). 

\subsection{Evaluation of the Controlled System}

In this section we evaluate the behavior of the system when the controller reacts to the system's evolution.
Results are reported for different clusters with the controller required to reach a set of degradation factors (\(\epsilon \in [0, 0.5]\)).
Experiments are repeated 30 times for each combination of cluster and degradation value.

\begin{figure}%
[t]%
     \centering
     \begin{subfigure}[c]{0.49\textwidth}
         \centering
         \includegraphics[width=\textwidth]{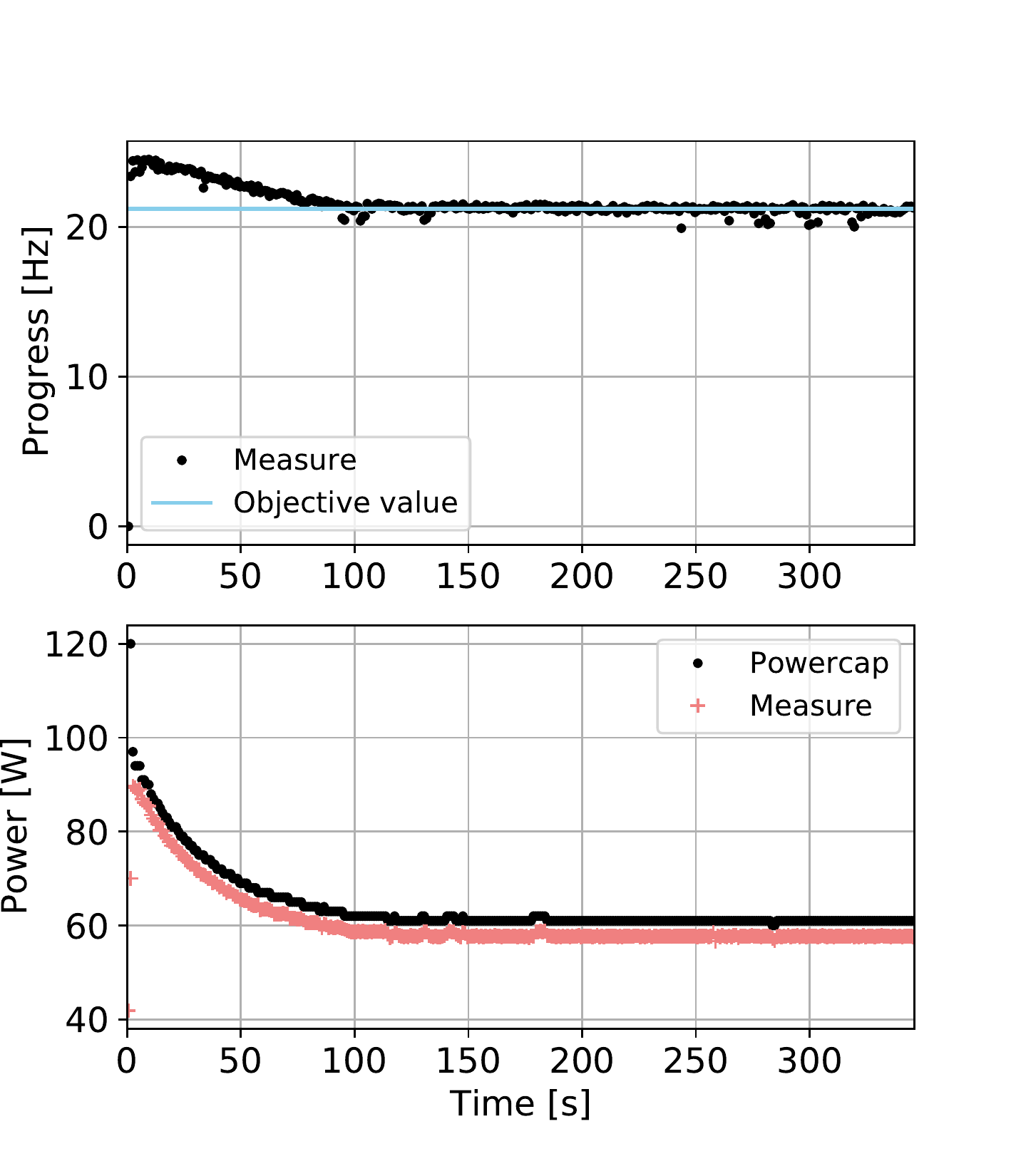}
         \caption{
         Controlled system behavior:
         Progress and powercap trough time
         (\(\epsilon=0.15\), \cluster{gros} cluster).
         Single representative execution.
         }
         \label{fig:control-time-85-gros}
     \end{subfigure}
     \hfill
     \begin{subfigure}[c]{0.49\textwidth}
         \centering
         \includegraphics[width=\textwidth,keepaspectratio,]{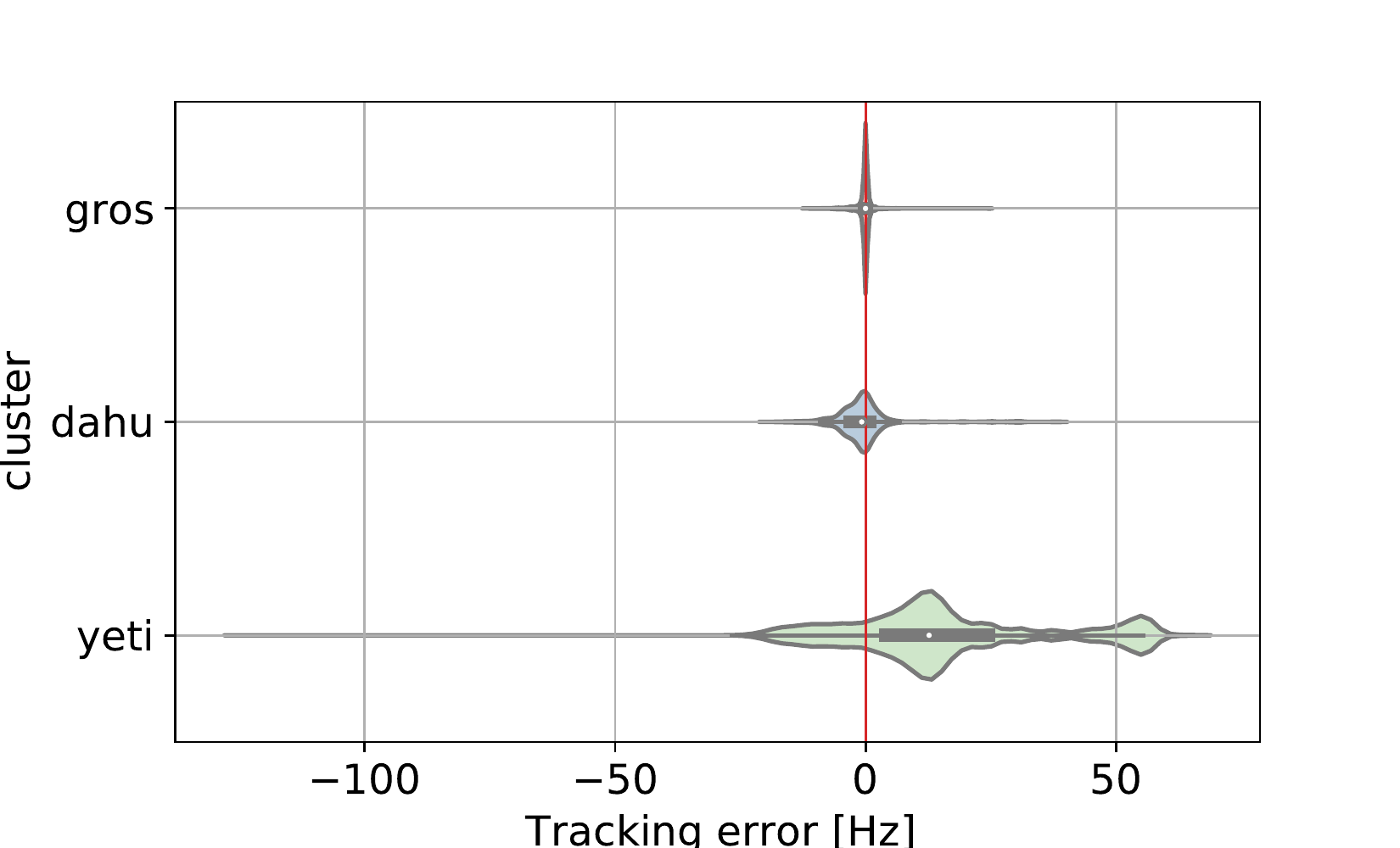}
	     \caption{
	     Distribution of the tracking error per cluster.
	     The sub-figure aggregates all executions run with the controller.
	     }
         \label{fig:controller-violin}
     \end{subfigure}
     \hfill
        \caption{Evaluation of the controlled system.}
        \label{fig:ctrl-evaluation}
\end{figure}

\Cref{fig:control-time-85-gros} shows a typical controlled system behavior through time on the \cluster{gros} cluster. The initial powercap is set at its upper limit, and the controller smoothly decreases its value until progress reaches the objective level (15\% degradation here). Note that the controlled system behavior shows neither oscillation nor degradation of the progress below the allowed value.
We report in \cref{fig:controller-violin} the distribution of the tracking error: the difference between the progress setpoint chosen by the controller and the measured progress.
The depicted distributions aggregate all experiments involving the controller.
The distributions of the error for the \cluster{gros} and \cluster{dahu} clusters are unimodal, centered near 0 (\(-0.21\) and \(-0.60\), \resp) 
with a narrow dispersion (\(1.8\) and \(6.1\), \resp). 
On the other hand, the distribution of the error for the \cluster{yeti} cluster exhibits two modes: the second mode (located between \SIrange[range-phrase={ and }]{50}{60}{\hertz}) is due to the model limitations.
For reasons to be investigated, the progress sometimes drops to about \SI{10}{\hertz} regardless of the requested power cap.
This behavior is notably visible in \cref{fig:identification-stairs-yeti} and is on a par with the observation that the more sockets a system has, the noisier it becomes.

\begin{figure}
[ht]
     \centering
     \begin{subfigure}[b]{0.3\textwidth}
         \centering
         \includegraphics[height=0.26\textheight]{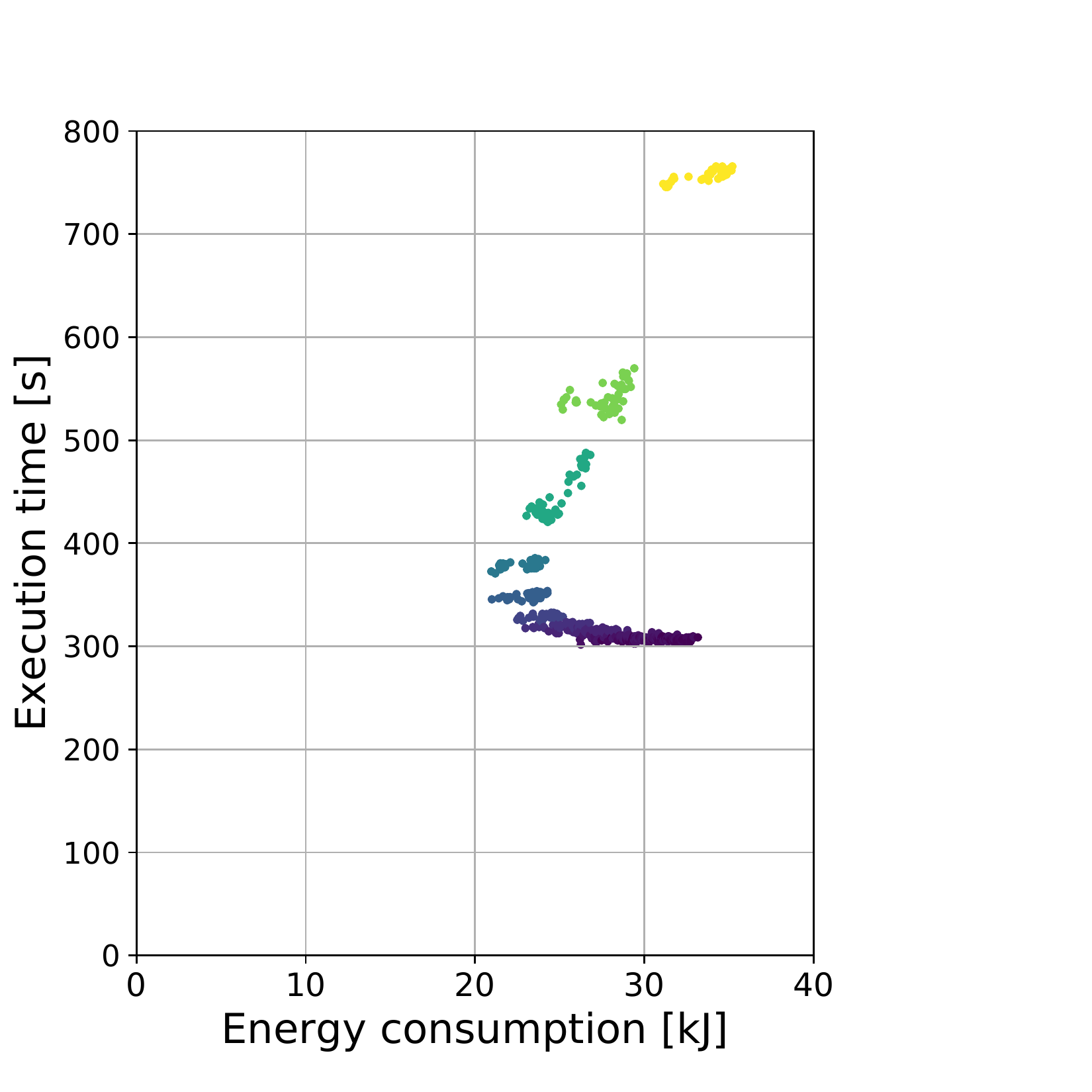}
         \caption{\cluster{gros} cluster}
         \label{fig:pareto-gros}
     \end{subfigure}
     \hfill
     \begin{subfigure}[b]{0.3\textwidth}
         \centering
         \includegraphics[height=0.26\textheight]{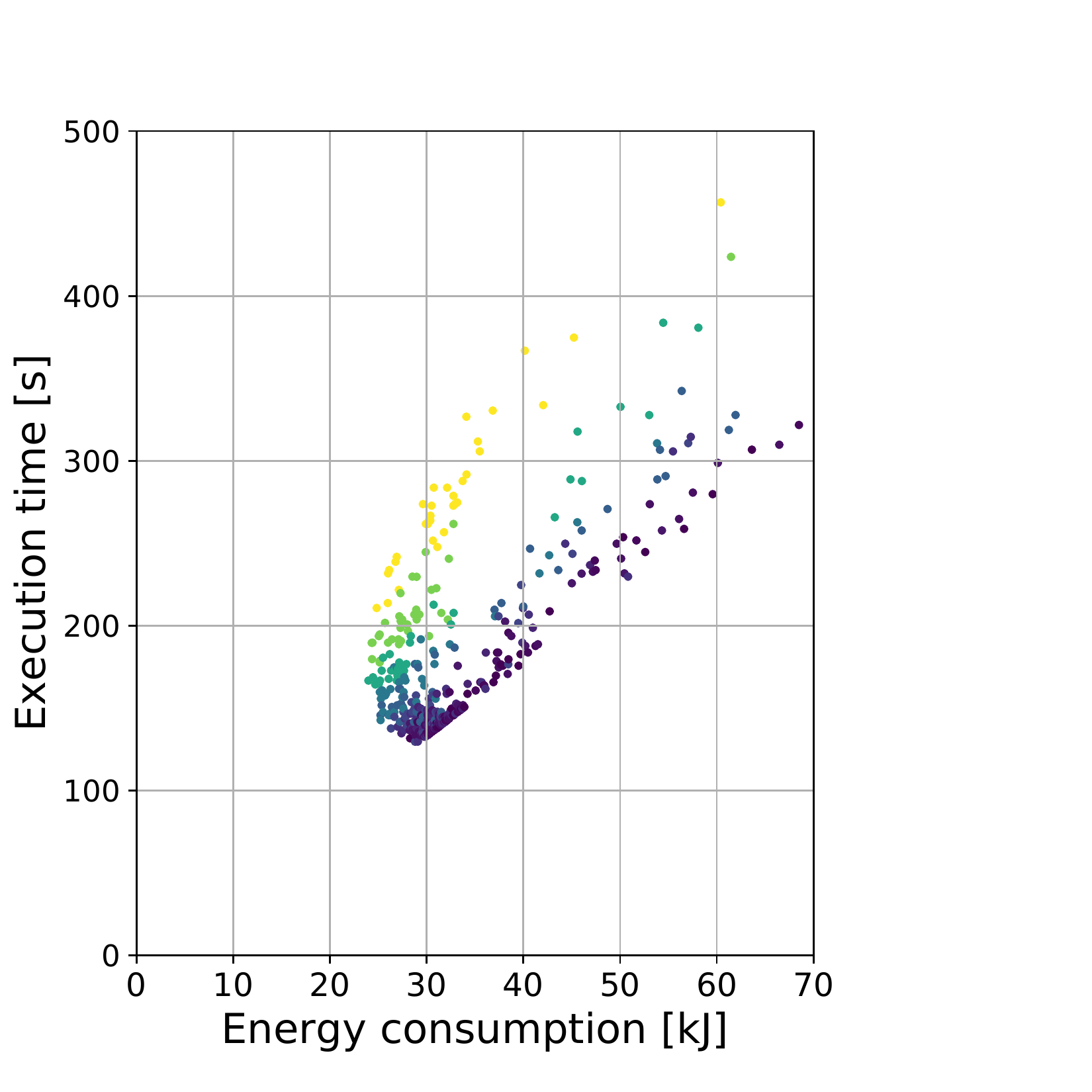}
         \caption{\cluster{dahu} cluster}
         \label{fig:pareto-dahu}
     \end{subfigure}
     \hfill
     \begin{subfigure}[b]{0.36\textwidth}
         \centering
         \includegraphics[height=0.26\textheight]{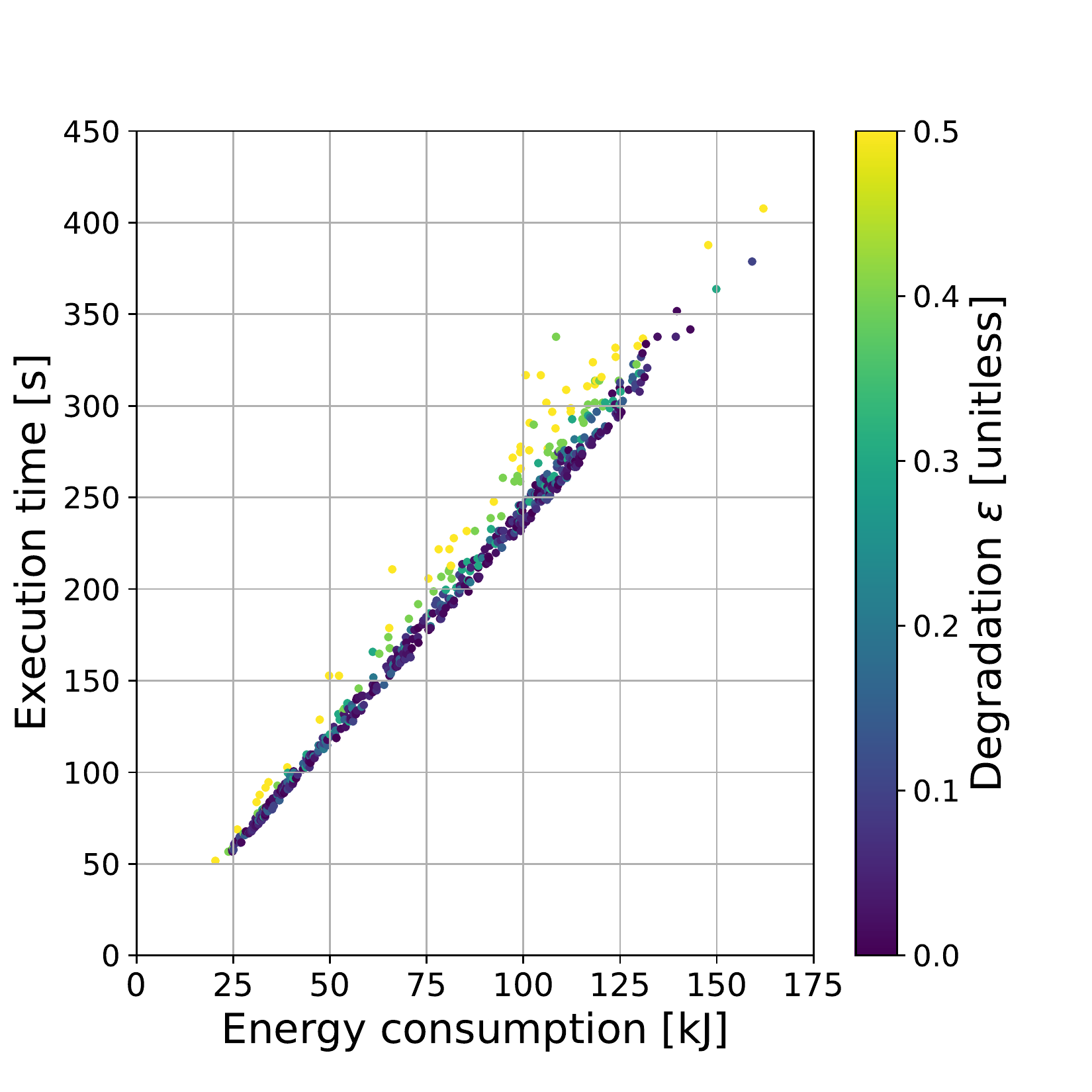}
         \caption{\cluster{yeti} cluster}
         \label{fig:pareto-yeti}
     \end{subfigure}
     \hfill
	\caption{
		Execution time with respect to energy consumption.
		Color indicates the requested degradation level \(\epsilon\).
		Each point depicts a single execution.
	}
	\label{fig:pareto}
\end{figure}

The controller objective is to adapt the benchmark speed by adapting the powercap on the fly, meaning we are exploiting \emph{time-local} behavior of the system.
Nevertheless, we are interested in the \emph{global} behavior of the benchmark, in other words, the total execution time and the total energy consumption.
To this end, we assessed the performance of the controlled system with a post mortem analysis.
We tested a total of twelve degradation levels ranging from \(0.01\) to \(0.5\), and ran each configuration a minimum of thirty times.
\Cref{fig:pareto} depicts the total execution time and the total energy consumption for each tested degradation level \(\epsilon\) in the time/energy space.
The experiments unveil a Pareto front for the \cluster{gros} and \cluster{dahu} clusters for degradation levels ranging from \SIrange{0}{15}{\percent} (\cref{fig:pareto-gros,fig:pareto-dahu}): this indicates the existence of a family of trade-offs to save energy.
For example, the \(\epsilon=0.1\) degradation level on the \cluster{gros} cluster is interesting because it allows, on average, saving \SI{22}{\percent} energy at the cost of a \SI{7}{\percent} execution time increase when compared with the baseline execution (\(\epsilon=0\) degradation level).
Note that degradation levels over \SI{15}{\percent} are not interesting because the increase in the execution time negates the energy savings.
The behavior on the \cluster{yeti} cluster is too noisy to identify interesting degradation levels.
However, the proposed controller does not negatively impact the performance.


\subsubsection{Discussion}
\label{ssec:discussion}

The proposed controller has an easily configured behavior: the user has to supply only an acceptable degradation level.
The controller relies on a simple model of the system, and it is stable on the \cluster{gros} and \cluster{dahu} clusters.
The approach does, however, show its limits on the \cluster{yeti} cluster, since the model is unable to explain the sporadic drops to \SI{10}{\hertz} in the application's progress.
This is particularly visible on \cref{fig:identification-stairs-yeti} between \SIrange[range-phrase={ and }]{60}{70}{\second}.
Nevertheless, we observe that these events can be characterized by the wider gap between the requested powercap and the measured power consumption.
We suspect that the number of packages and the NUMA architecture, or exogenous temperature events, are responsible for these deviations.
Further investigations are needed to confirm this hypothesis. If it is confirmed, development of control strategies will be considered for integrating distributed actuation or temperature disturbance anticipation.

\paragraph{On the generalization to other benchmarks.}

The use of STREAM benchmark is motivated by its stable memory-bound profile, and its ability to be instrumented with heartbeats---thanks to its straightforward design.
\Cref{fig:identification-stairs-model,fig:identification-stairs} illustrate this stability: while no powercap change occurs, performance is largely constant.
The presented approach extends similarly for memory intensive phases of applications.
We expect compute-bound phases to show a different (simpler) power to progress profile (\cref{fig:sc-model}), less amenable to optimization.
Indeed, every power increase should improve performance, even for high powercap values.
Furthermore, with such linear behavior, modeling will have to be recomputed, and controller parameters updated.
Overall, controlling an application with varying resource usage patterns thus requires \emph{adaptation}---a control technique implying automatic tuning of the controller parameters---to handle powercap-to-progress behavior transitions between phases.
It is a natural direction of research for future work.

\section{Related Work}
\label{sec:related-works}

\paragraph{On power regulation in HPC.}

A large body of related work seeks to optimize performance or control
energy consumption on HPC systems using a wide range of control
knobs\cite{BhalachandraS2015Using,DutotPF2017Towards,OrgerieAC2008Save,PetoumenosP2015Power}. Most of
these methods, however, target a different objective from our work or are based either on static schemes used at the beginning of
a job or on simple loops without formal performance guarantees. This is also the
case for GeoPM\cite{EastepJ2017Global}, the most prominent available power management infrastructure
for HPC systems. An open source framework designed by Intel, GeoPM uses the same actuator than
as our infrastructure (RAPL) does but with application-oblivious monitoring (PMPI or OMPT) capabilities.

\paragraph{On using control theory for power regulation.}

Control-theory-based approaches to power regulation focus mainly on web servers\cite{AbdelzaherT2008Introduction}, clouds\cite{ZhouY2016CASH}, and real-time systems\cite{ImesC2015POET} applications. They typically leverage DVFS\cite{AlbersS2011Algorithms} as an actuator and formulate their objectives in terms of latency. Controllers are adaptive, with an update mechanism to cope with unmodeled external disturbances. The present work stands out for  two reasons. First, it uses Intel's RAPL mechanism, a unified architecture-agnostic and future-proof solution, to leverage power. Moreover, we do not consider applications with predefined latency objectives but instead focus on the \emph{science} performed by HPC applications, based on a heartrate progress metric.
Other works similarly use RAPL in web servers\cite{LoD2014Towards} and real-time systems\cite{ImesC2019CoPPer} contexts and non-latency-based performance metrics\cite{SantriajiMH2016GRAPE}. To the best of our knowledge, however, we present the first control theory approach to power regulation using RAPL for HPC systems.






\section{Conclusion}
\label{sec:conclusion}

We address the problem of managing energy consumption in complex heterogeneous HPC by focusing on the potential of
dynamically adjusting power across compute elements to save energy with limited and controllable impact on performance.
Our approach involves using control theory, with a method adapted to HPC systems,
and leads to identification and controller design for the targeted system.
Experimental validation shows good results for systems with lower numbers of sockets running a memory-bound benchmark.
We identify limiting aspects of our method and
areas for further study, such as integrating measures of the temperature or extending the control to heterogeneous
devices.

\subsubsection*{Acknowledgments and Data Availability Statement}

Experiments presented in this paper were carried out using the Grid'5000 testbed, supported by a scientific interest group hosted by Inria and including CNRS, RENATER and several Universities as well as other organizations (see \url{https://www.grid5000.fr}).
Argonne National Laboratory's work was supported by the U.S.\@ Department of
Energy, Office of Science, Advanced Scientific Computer Research, under
Contract DE-AC02-06CH11357. This research was supported by the Exascale
Computing Project (17-SC-20-SC), a collaborative effort of the U.S.\@ Department of Energy Office
of Science and the National Nuclear Security Administration.
This research is partially supported by the NCSA-Inria-ANL-BSC-JSC-Riken-UTK Joint-Laboratory for Extreme Scale Computing (JLESC, \url{https://jlesc.github.io/}).

The datasets and code generated and analyzed during the current study are available in the Figshare repository: \url{https://doi.org/10.6084/m9.figshare.14754468}\cite{artifacts}.

%
%
%
\bibliographystyle{splncs04}
\bibliography{biblio}

\end{document}